\documentclass[conference,a4paper]{IEEEtran}

\usepackage{cite}
\usepackage{amsmath,amssymb,amsfonts}
\usepackage{graphicx}
\usepackage{textcomp}
\usepackage{xcolor}
\usepackage{algorithm}
\usepackage{algpseudocode}
\usepackage{braket}
\usepackage[capitalize]{cleveref}
\usepackage{mathtools}
\IEEEoverridecommandlockouts

\newcommand\eqc{\stackrel{\text{c}}{=}}
\makeatletter
\newcommand{\linebreakand}{%
  \end{@IEEEauthorhalign}
  \hfill\mbox{}\par
  \mbox{}\hfill\begin{@IEEEauthorhalign}
}
\makeatother

\usepackage{tikz}

\newcommand\copyrighttext{%
  \footnotesize \textcopyright \the\year{} IEEE. Personal use of this material is permitted. Permission from IEEE must be obtained for all other uses, including reprinting/republishing this material for advertising or promotional purposes, collecting new collected works for resale or redistribution to servers or lists, or reuse of any copyrighted component of this work in other works.}

\newcommand\copyrightnotice{%
\begin{tikzpicture}[remember picture,overlay]
\node[anchor=south,yshift=10pt] at (current page.south) {\fbox{\parbox{\dimexpr0.75\textwidth-\fboxsep-\fboxrule\relax}{\copyrighttext}}};
\end{tikzpicture}%
}

\begin{document}

\title{Variational Inference for Quantum HyperNetworks}

% \author{\IEEEauthorblockN{Luca Nepote}
% \textit{EURECOM}\\
% Sophia Antipolis, France \\
% luca.nepote@eurecom.fr}
% \and
% \IEEEauthorblockN{Alix Lhéritier}
% \IEEEauthorblockA{\textit{Research Department} \\
% \textit{Amadeus}\\
% Sophia Antipolis, France \\
% alix.lheritier@amadeus.com}
% \and
% \IEEEauthorblockN{Nicolas Bondoux}
% \IEEEauthorblockA{\textit{Research Department} \\
% \textit{Amadeus}\\
% Sophia Antipolis, France \\
% nbondoux@amadeus.com}
% \linebreakand
% \IEEEauthorblockN{Marios Kountouris}
% \and
% \IEEEauthorblockN{Maurizio Filippone}
% \IEEEauthorblockA{\textit{Statistics Program} \\
% \textit{KAUST}\\
% Thuwal, Saudi Arabia \\
% maurizio.filippone@kaust.edu.sa}
% }

\author{\IEEEauthorblockN{Luca Nepote\IEEEauthorrefmark{1}\IEEEauthorrefmark{2}, Alix Lhéritier\IEEEauthorrefmark{1}, Nicolas Bondoux\IEEEauthorrefmark{1}, Marios Kountouris\IEEEauthorrefmark{2}\IEEEauthorrefmark{3} and Maurizio Filippone\IEEEauthorrefmark{4}}
\IEEEauthorblockA{\IEEEauthorrefmark{1}Amadeus, Sophia Antipolis, France}
\IEEEauthorblockA{\IEEEauthorrefmark{2}EURECOM, Sophia Antipolis, France}
\IEEEauthorblockA{\IEEEauthorrefmark{3}Universidad de Granada, Spain}
\IEEEauthorblockA{\IEEEauthorrefmark{4}KAUST, Statistics Program, Saudi Arabia}
\thanks{Correspondence to: Luca Nepote \textless luca.nepote@eurecom.fr\textgreater, Alix Lhéritier \textless alix.lheritier@amadeus.com\textgreater.}
}

\maketitle
\copyrightnotice

\begin{abstract}
Binary Neural Networks (BiNNs), which employ single-bit precision weights, have emerged as a promising solution to reduce memory usage and power consumption while maintaining competitive performance in large-scale systems. However, training BiNNs remains a significant challenge due to the limitations of conventional training algorithms. Quantum HyperNetworks offer a novel paradigm for enhancing the optimization of BiNN by leveraging quantum computing. Specifically, a Variational Quantum Algorithm is employed to generate binary weights through quantum circuit measurements, while key quantum phenomena such as superposition and entanglement facilitate the exploration of a broader solution space. In this work, we establish a connection between this approach and Bayesian inference by deriving the Evidence Lower Bound (ELBO), when direct access to the output distribution is available (i.e., in simulations), and introducing a surrogate ELBO based on the Maximum Mean Discrepancy (MMD) metric for  scenarios involving implicit distributions, as commonly encountered in practice. Our experimental results demonstrate that the proposed methods outperform standard Maximum Likelihood Estimation (MLE), improving trainability and generalization.
\end{abstract}


\section{Introduction}
Machine Learning (ML) is transforming industries and shaping daily life; however, the substantial computational power required to train and deploy advanced models leads to unsustainable energy consumption \cite{patterson,strubell2020energy}. As ML models continue to increase in complexity and scale, improving their energy efficiency has become a critical and urgent challenge.  

To address this challenge, several techniques have been proposed, with quantization standing out as a particularly effective approach \cite{yuan2023comprehensive}. By reducing the precision of each model weight, quantization lowers memory usage requirements and speeds up computation, while maintaining competitive performance \cite{wang2025optimizing}. The most extreme form of quantization is \textit{binarization}, where each parameter is constrained to a single bit representing values in $\{-1,+1\}$. \textit{Binary Neural Networks} (BiNNs) \cite{qin2020binary,courbariaux1602binarynet} leverage this principle to drastically reduce energy consumption and memory usage. When scaled appropriately, BiNNs can achieve performance comparable to their floating-point (full-precision) counterparts \cite{wang2023bitnet}, making them a promising solution for efficient deep learning models at scale.  

However, the discrete nature of binary weights poses significant challenges for optimization. Traditional deep learning training techniques, which rely on continuous weight updates, are not well-suited to the constraints imposed by BiNNs. As a result, alternative training strategies are necessary to overcome these limitations, requiring the development of novel optimization methods for binary-weight networks \cite{yin2018understanding}.  
A promising approach for training such models lies in the use of Quantum Computing. When a quantum circuit is measured, the resulting bitstrings can be directly employed as binary weights in a BiNN, effectively addressing the limitations of single-bit precision and eliminating the need for backpropagation. Furthermore, by leveraging fundamental quantum phenomena such as \textit{superposition} and \textit{entanglement}, it becomes possible to explore a larger solution search space in quantum superposition, potentially leading to more efficient and effective optimization. 

This work investigates and advances a promising approach known as \textit{Quantum HyperNetworks} \cite{carrasquilla2023quantum}: a training technique at the intersection of quantum computing and machine learning. Specifically, we focus on \textit{Quantum Machine Learning} (QML) and \textit{Variational Quantum Algorithms} (VQAs) \cite{cerezo2021variational} to explore how quantum computing can be integrated into deep learning systems for more efficient and scalable training.

As highlighted in \cite{carrasquilla2023quantum}, Quantum HyperNetworks are inherently linked to the Bayesian approach, as their output defines a probability distribution over the weights of a BiNN. The core contribution of our work is the derivation of the Evidence Lower Bound (ELBO) for Quantum HyperNetworks, along with a surrogate version tailored for practical implementation. These formulations introduce an implicit form of regularization, thereby enhancing both trainability and generalization. While prior studies \cite{benedetti2021variational,liu2018differentiable} have already studied the use of distance metrics such as Kullback-Leibler (KL) divergence and the Maximum Mean Discrepancy (MMD) in quantum generative modeling - focusing on data generation and the representation power of quantum circuits - our work applies variational inference \cite{blei2017variational} to derive regularization terms based on KL and MMD within the training loss. We analyze the influence of these terms on the optimization process, highlighting their role in enhancing performance and stability.

\section{Preliminaries}
As previously mentioned, \textit{Binary Neural Networks} offer a promising solution for both energy- and memory-limited devices. However, a major challenge in training these models lies in the nested optimization loop required to update parameters, hyperparameters, and architectural choices. This process is computationally intensive due to the presence of both inner and outer optimization loops, resulting in a significant computational cost and an unsustainable carbon footprint, similar to their floating-point counterparts. Indeed, even with BiNNs, floating-point numbers are required for the parameter updates during training \cite{courbariaux1602binarynet}.

In \cite{carrasquilla2023quantum}, the authors introduced \textit{Quantum HyperNetworks} as a novel approach to address this challenge. A \textit{HyperNetwork} \cite{ha2017hypernetworks} is a machine learning technique in which one network generates the weights for another, typically larger, network, accelerating the search within high-dimensional parameter spaces containing millions of variables \cite{Schmidhuber1992LearningTC}.
This study explored a quantum-inspired strategy that jointly optimizes all the \textit{augmented parameters}---i.e. parameters, hyperparameters, and architectural choices---within a single optimization loop by leveraging quantum states. By exploiting key quantum properties such as superposition and entanglement, the approach enables more efficient exploration of the solution space, reducing computational overhead.

\subsection{VQA for Quantum HyperNetworks}
Our objective is to identify ideally optimal solutions for the classical supervised ML problem, based on a Neural Network function $\text{NN}(\mathbf{x}; \mathbf{w})$, with input $\mathbf{x}$ and augmented parameters $\mathbf{w}\in\{-1,+1\}^N$ with the following cost function
\begin{equation}
C(\mathbf{\boldsymbol{\sigma}}) = \frac{1}{N_s} 
\sum_{i=1}^{N_s} \ell \big( \text{NN}(\mathbf{x}_i; 2\boldsymbol{\sigma}-1), \mathbf{y}_i \big)
\end{equation}
where $N_s$ denotes the training set size, $\mathbf{x}_i$  represents an input sample with the corresponding label $\mathbf{y}_i$ in the dataset $(Y;X)$, with \( X = \{ \mathbf{x}_i \}_{i=1}^{N_s} \) and \( Y = \{ \mathbf{y}_i \}_{i=1}^{N_s} \), and $\ell$ is the loss function for the supervised ML problem.
To encode the problem into a quantum state suitable for a VQA, each parameter is mapped onto one of $N$ qubits in the computational basis,
\begin{equation}
|\Psi\rangle = \sum_{\sigma_1, \ldots, \sigma_N} \Psi(\sigma_1, \ldots, \sigma_N) |\sigma_1, \ldots, \sigma_N\rangle,
\end{equation}
with $\sigma_i \in \{0,1\}$. One way to build a Hamiltonian from the classical cost function is to consider the Pauli operator $Z$ acting on the $i$-th qubit, i.e.
\begin{equation}
     \hat{\sigma}_i^z \equiv I^{\otimes i-1} \otimes Z \otimes I ^{\otimes n-i},
\end{equation}
with the eigenvalue equation
\begin{equation}
\hat{\sigma}_i^z |\sigma_1, \ldots, \sigma_i ,\ldots , \sigma_N\rangle = \left( 2 \sigma_i  -1 \right)|\sigma_1, \ldots, \sigma_i ,\ldots , \sigma_N\rangle.
\end{equation}
%Note that the eigenvalues are in $\{-1,+1\}$ as the weights and biases of the NN.
Let us consider $w_i\equiv 2 \sigma_i  -1$.
By noting that any Boolean function $C(\sigma_1,\dots,\sigma_N)$ can be rewritten as a multilinear combination of the variables $w_i$ and applying the same multilinear combination to the $\hat{\sigma}_i^z$ operators, the following $2^N \times 2^N$ diagonal Hamiltonian is obtained
\begin{equation}
    \hat{C} = 
    \begin{pmatrix}
        C(\boldsymbol{\sigma}_1) & 0 & \cdots & 0 \\
        0 & C(\boldsymbol{\sigma}_2) & \cdots & 0 \\
        \vdots & \vdots & \ddots & \vdots \\
        0 & 0 & \cdots & C(\boldsymbol{\sigma}_{2^N})
    \end{pmatrix} ,
    \label{cost operator}
\end{equation}
where each diagonal element of this operator represents the cost value associated with a specific BiNN configuration $\boldsymbol{\sigma}_i$, resulting in a total of $2^N$ possible BiNNs. Notably, other encodings are possible, enabling the use of alternative computational bases to suit different optimization strategies.

Each basis element corresponds to a specific configuration of augmented parameters, encompassing all model parameters, hyperparameters, and architectural choices of the BiNN. Any additional decision is encoded using an extra qubit: for instance, if the choice involves selecting between two activation functions $a_1$ and $a_2$, an additional qubit $\sigma$ is introduced, and the activation is determined based on its value
\begin{equation}
a(\mathbf{x}; \sigma) =
\begin{cases} 
a_1(\mathbf{x}) & \text{if } \sigma = 0, \\
a_2(\mathbf{x}) & \text{if } \sigma = 1.
\end{cases}
\end{equation}
It is important to note that, unlike the previous discussion on BiNNs, activations in this approach are not restricted to binary values---only the weights are binary.

When the quantum state is measured, it yields the augmented parameters for the BiNN. This encoding is well-suited for a VQA, where, after training, the system is expected to return an optimized Neural Network (NN) configuration.

The quantum state $|\Psi\rangle$ is obtained by a parametrized quantum circuit $U(\boldsymbol{\theta})$, such that 
\begin{equation}
    |\Psi_{\boldsymbol{\theta}}\rangle = U(\boldsymbol{\theta}) |0\rangle^{\otimes n}.
\end{equation}

The quantum circuit is constructed as a sequence of $L$ unitary blocks, that is, $U(\mathbf{\boldsymbol{\theta}}) = U_L(\mathbf{\boldsymbol{\theta}}_L) \dots U_1(\mathbf{\boldsymbol{\theta}}_1)$, where each block consists of a set of linear operations, namely \textit{CX}, \textit{RY}, and \textit{RZ} gates. The circuit depth is adjustable, allowing control over the degree of entanglement within the quantum system. Each set of unitary operations is expressed as
\begin{align}
    U_k(\boldsymbol{\theta}_k) &= \prod\limits_{\substack{m=1+k \bmod 2 \\ \text{step} \, 2}}^{N-1} 
    \text{CX}(m, m+1) \notag \\
    &\quad \times \prod\limits_{j=1}^{N} \text{RY}(j, \theta_{0,j,k}) \text{RZ}(j, \theta_{1,j,k}).
    \label{eq:Uk_theta}
\end{align}

The observable is defined through a stochastic relaxation of the discrete optimization problem, enabling the search for an optimal solution in the augmented parameter distribution. Mathematically, the expectation value of the Hamiltonian $\hat{C}$---the energy $E$---is defined as:
\begin{equation}
    E(\boldsymbol{\theta}) = \langle \Psi_{\boldsymbol{\theta}}|\hat{C}|\Psi_{\boldsymbol{\theta}}\rangle 
\end{equation}
and the best approximation to the problem is given by 
\begin{equation}
    \boldsymbol{\theta}^* = \arg \min_{\boldsymbol{\theta}} E(\boldsymbol{\theta}),
\end{equation}
since $\min_{\boldsymbol{\theta}} E(\boldsymbol{\theta})\geq \lambda^\star$, $\lambda^\star$ being the lowest eigenvalue of $\hat{C}$, with equality if there exists $\boldsymbol{\theta}^\star$ such that $\ket{\Psi_{\boldsymbol{\theta}^\star}}$ is equal to the eigenvector corresponding to $\lambda^\star$.

\subsection{Optimization}
For the optimization, the standard gradient descent method is used in \cite{carrasquilla2023quantum}. The expected value $E(\boldsymbol{\theta})$ and the gradient $\nabla_{\boldsymbol{\theta}}E(\boldsymbol{\theta})$ are estimated through statistical averages over multiple quantum circuit measurements. These values are then processed by a classical optimizer to iteratively update the parameters until convergence. Consider
\begin{align}
    E(\boldsymbol{\theta}) 
    &= \langle \Psi_{\boldsymbol{\theta}} | \hat{C} | \Psi_{\boldsymbol{\theta}} \rangle \notag\\
    &= \sum_{\sigma_1, \sigma_2, \ldots, \sigma_N} |\Psi_{\boldsymbol{\theta}}(\sigma_1, \sigma_2, \ldots, \sigma_N)|^2 C(\sigma_1, \sigma_2, \ldots, \sigma_N) \notag \\
    &= \mathbb{E}_{\tilde{\boldsymbol{\sigma}} \sim |\Psi_{\boldsymbol{\theta}}|^2} [C(\tilde{\boldsymbol{\sigma}})] \approx \frac{1}{N_{qc}} \sum_{i=1}^{N_{qc}} C(\tilde{\boldsymbol{\sigma}}_i)
    \label{eq:expectation}
\end{align}
where $\tilde{\boldsymbol{\sigma}_i}$ is  one of the $N_{qc}$ sampled BiNN configurations from the quantum circuit. Considering this Monte-Carlo method is crucial because, if an exhaustive search of all possible BiNN configurations was conducted to construct the Hamiltonian, the optimal solution would already be known before running the VQA - making the quantum algorithm unnecessary \cite{naim2024scalable}.

In a real experimental setting, the \textit{ parameter shift rule} \cite{mitarai2018quantum,schuld2019evaluating} is used to compute gradients efficiently:
\begin{equation}
    \frac{\partial E(\boldsymbol{\theta})}{\partial \theta_{\alpha,j,k}} = \frac{1}{2} \left[ E(\boldsymbol{\theta}_{\alpha,j,k}^+) - E(\boldsymbol{\theta}_{\alpha,j,k}^-) \right]
    \label{ps}
\end{equation}
where the elements of the shifted parameter vector $\boldsymbol{\theta}_{\alpha,j,k}^{\pm}$ are
\begin{equation}
    \theta_{\beta,m,l}^{\pm} = \theta_{\beta,m,l} \pm \frac{\pi}{2} \delta_{\alpha,\beta}\delta_{m,j}\delta_{k,l}.
    \label{eq: ps for each param}
\end{equation}
The expected values for the gradient are estimated by a Monte Carlo approach for each shifted parameter.
Previous work \cite{carrasquilla2023quantum} uses a \textit{Tensor Network}\cite{biamonte2017tensor} simulator combined with automatic differentiation, employing the L-BFGS algorithm for optimization.

%%%%%%%%%%%%%%%%%%%%%%%%%%%%%%%%%%%%%%%%%%%%%%%%%%%%%%%%%%%%%%%%%%%%%%%%%%%%%%%%%
\section{Proposed Methodology}
In our implementation, we design a circuit with linear connectivity following \eqref{eq:Uk_theta}, and perform measurements in the Pauli Z basis. We initialize the system in the zero state vector but instead of directly constructing the Hamiltonian of \eqref{cost operator}, we employ \eqref{eq:expectation} with the parameter shift rule of \eqref{ps}. This ensures compliance with practical quantum settings while maintaining computational efficiency. 

As noted by the authors in \cite{carrasquilla2023quantum}, the Quantum HyperNetworks approach established a direct connection to Bayesian inference. Specifically, the quantum circuit $\ket{\Psi_{\boldsymbol{\theta}}}$ defines a probability distribution over the binary weights of the BiNN. While direct evaluation of the posterior distribution is intractable, we can approximate it using Variational Inference (VI) by estimating the ELBO.
We consider here two cases: (i) we assume full access to the quantum circuit distribution, such as in simulations where the amplitudes and corresponding probabilities of each bitstring are known; (ii) we adopt a more realistic setting in which the quantum circuit distribution is implicit and only accessible through samples (i.e., quantum measurements).

\subsection{Explicit ELBO}
Following a VI approach, we can approximate the true posterior distribution over binary weights using a parametric family $q_{\boldsymbol{\theta}}(\boldsymbol{\sigma})$, where $\boldsymbol{\theta}$ represents the variational parameters. In our case, $\boldsymbol{\theta}$ corresponds to the rotational angles of the $RZ$ and $RY$ gates in the quantum circuit. The probability distribution over the binary weights is governed by the \textit{Born rule}
\begin{equation}
    q_{\boldsymbol{\theta}}(\boldsymbol{\sigma}) = | \bra{\boldsymbol{\sigma}}  {U(\mathbf{\mathbf{\boldsymbol{\theta}}})}\ket{\mathbf{0}}|^2,
\end{equation}
which defines the probability of measuring each bitstring as a function of the variational parameters $\boldsymbol{\theta}$.

Considering the general ELBO expression
\begin{equation}
    \mathcal{L}_{\text{ELBO}} = \mathbb{E}_{q_{\boldsymbol{\theta}}(\boldsymbol{\sigma})} \left[ \log p(Y \mid X, \boldsymbol{\sigma}) \right] - \mathrm{KL}[q_{\boldsymbol{\theta}}(\boldsymbol{\sigma}) \| p(\boldsymbol{\sigma})],
\label{ELBO}
\end{equation}
the first term is the model fitting term (expected log-likelihood), which we approximate using Monte Carlo sampling by measuring the quantum circuit. The estimator is unbiased, with variance scaling as $1/N_{\text{qc}}$, where $N_{\text{qc}}$ denotes the number of measurements. In the context of a classification problem, $X$ corresponds to the set of data points with labels $Y$. The second term serves as a penalty or regularization term, corresponding to the KL divergence between the variational distribution and the prior distribution over the weights of the BiNN. 
For this derivation, we impose a uniform prior for each parameter $\{0,1\}$:
\begin{equation}
\sigma_i \sim \mathcal{U}\{0,1\},
\end{equation}
with $\sigma_i$ being the $i$-th component of the bitstring $\boldsymbol{\sigma}$.

Extending this to all binary weights, we obtain a joint uniform distribution over all possible BiNN configurations. This implies that each configuration has an equal probability of being sampled, specifically $1/N_\text{t}$, where $N_\text{t} = 2^{n_{\text{qubits}}}$ represents the total number of possible configurations.

Given the uniform prior $p$, we derive the following expression for the regularization term in \eqref{ELBO}:
\begin{align}
    \mathrm{KL}[q_{\boldsymbol{\theta}}(\boldsymbol{\sigma}) \| p(\boldsymbol{\sigma})] &= \sum_{\boldsymbol{\sigma}} q_{\boldsymbol{\theta}}(\boldsymbol{\sigma}) \log \frac{q_{\boldsymbol{\theta}}(\boldsymbol{\sigma})}{p(\boldsymbol{\sigma})} \notag \\
    &= \sum_{\boldsymbol{\sigma}} q_{\boldsymbol{\theta}}(\boldsymbol{\sigma}) \log q_{\boldsymbol{\theta}}(\boldsymbol{\sigma}) - \sum_{\boldsymbol{\sigma}} q_{\boldsymbol{\theta}}(\boldsymbol{\sigma}) \log p(\boldsymbol{\sigma}) \notag \\
    &= \sum_{\boldsymbol{\sigma}} q_{\boldsymbol{\theta}}(\boldsymbol{\sigma}) \log q_{\boldsymbol{\theta}}(\boldsymbol{\sigma}) - \sum_{\boldsymbol{\sigma}} q_{\boldsymbol{\theta}}(\boldsymbol{\sigma}) \log \frac{1}{N_\text{t}} \notag \\
    &= \sum_{\boldsymbol{\sigma}} q_{\boldsymbol{\theta}}(\boldsymbol{\sigma}) \log q_{\boldsymbol{\theta}}(\boldsymbol{\sigma}) + \log N_\text{t},
\end{align}
where the first term corresponds to the negative entropy of the quantum circuit variational distribution, and the second term is a constant that does not influence the optimization process.

By substituting this term into \eqref{ELBO}, we obtain the final expression to be maximized 
\begin{equation}
\mathcal{L}_{\text{ELBO}}  \eqc \mathbb{E}_{q_{\boldsymbol{\theta}}(\boldsymbol{\sigma})} \left[ \log p(Y \mid X, \boldsymbol{\sigma}) \right] %\notag %\\
%\quad 
- \mathbb{E}_{q_{\boldsymbol{\theta}}(\mathbf{\boldsymbol{\sigma}})} \left[ \log q_{\boldsymbol{\theta}}(\boldsymbol{\sigma}) \right],
\label{ELBO_approximate}
\end{equation}
where $\eqc$ denotes equality up to a constant term. Note that the second term of \eqref{ELBO_approximate} is the Shannon entropy of $q_{\boldsymbol{\theta}}$.

In experimental settings, accurately estimating the entropy of the output distribution requires collecting multiple samples, which can be computationally expensive. In contrast, simulations may provide direct access to the full state vector, allowing precise computation of the probability distribution over bitstrings.
% In this second case, we employ the Shannon entropy, defined as  
% \begin{equation}
%     \mathcal{H}_p \coloneqq  - \mathbb{E}_{q_{\boldsymbol{\theta}}(\mathbf{\boldsymbol{\sigma}})} \left[ \log q_{\boldsymbol{\theta}}(\boldsymbol{\sigma}) \right],
% \end{equation}  
% where $p(\boldsymbol{\sigma})$ corresponds to the output probability. This term can be incorporated into the ELBO formulation, yielding the final expression
% \begin{equation}
% \mathcal{L}_{\text{ELBO}} \eqc \mathbb{E}_{q_{\boldsymbol{\theta}}(\boldsymbol{\sigma})} \left[ \log p(Y \mid X, \boldsymbol{\sigma}) \right] + \mathcal{H}_{q_{\boldsymbol{\theta}}}.
% \label{ELBO final}
% \end{equation}

\subsection{Surrogate ELBO}
In this subsection, we derive a \textit{Surrogate Evidence Lower Bound} (SELBO) for implicit distributions. This approach is applicable to both simulations and real quantum hardware, where direct access to the quantum state distribution is unavailable, and only sampled data can be used. The challenge of applying variational inference to implicit distributions has also been explored in \cite{molchanov2019doublysemiimplicitvariationalinference}, where the authors consider semi-implicit distributions, constructed as a mixture of explicit conditional distributions.

We adopt the approach proposed in \cite{Pomponi_2021}, where the KL divergence regularization term is replaced with MMD \cite{JMLR:v13:gretton12a}. MMD is an integral probability metric \cite{Müller_1997} that compares two probability distributions using sample data rather than their explicit distributions. In our setting, one of the distributions is the implicit distribution $q_{\boldsymbol{\theta}}({\boldsymbol{\sigma}})$ over the binary weights, defined by the quantum circuit and accessible only through measurements. The other is the uniform prior $p(\boldsymbol{\sigma})$, from which we assume to have only a finite set of samples during computation. 
Then, given two sets of samples $\{x_i\}_{i=1}^{n} \sim q_{\boldsymbol{\theta}}(\boldsymbol{\sigma})$ and $\{y_j\}_{j=1}^{m} \sim p(\boldsymbol{\sigma})$, %the squared MMD is defined as:  
\begin{align}
&\operatorname{MMD}^2(q_{\boldsymbol{\theta}}(\boldsymbol{\sigma}), p(\boldsymbol{\sigma}))  \coloneqq \\ &\mathbb{E}_{x, x' \sim q_{\boldsymbol{\theta}}} \left[ k(x, x') \right] %\notag \\
 + \mathbb{E}_{y, y' \sim p} \left[ k(y, y') \right] \nonumber %\\
 - 2 \mathbb{E}_{\substack{x \sim q_{\boldsymbol{\theta}}\\ y \sim p}} \left[ k(x, y) \right], \notag
%\label{eq:MMD_modified}
\notag
\end{align} 
where $k(\cdot,\cdot)$ is a positive-definite kernel, which we assume to be the Gaussian radial basis function (RBF) kernel, i.e.%This is given by
\begin{equation}
k(x,y) = \exp\left(-\frac{\|x-y\|^2}{h^2}\right),
\end{equation}
with $h$ being the kernel bandwidth parameter. 
This parameter determines the scale of the Gaussian kernel: a small $h$ results in a narrow, localized kernel that closely approximates a Dirac delta function. This enhances sensitivity to local differences (low bias) between samples of the two distributions, but makes the estimation more susceptible to noise (high variance). Conversely, a large $h$ produces a broader, smoother kernel that reduces sensitivity to local variations (low variance) while emphasizing global differences. However, this comes at the cost/risk of higher bias, potentially overlooking important or finer details. Thus, choosing an appropriate $h$ is essential to effectively compare distributions, ensuring a balance between capturing local variations and preserving global structures.

We consider the squared MMD and the following  unbiased estimator \cite{JMLR:v13:gretton12a}
\begin{align}
\widehat{\mathrm{MMD}^2}_\text{U} &= \frac{1}{n(n-1)} \sum_{\substack{i,j=1 \\ i \neq j}}^{n} k(x_i, x_j) \notag\\
              &\quad+ \frac{1}{m(m-1)} \sum_{\substack{i,j=1 \\ i \neq j}}^{m} k(y_i, y_j) \nonumber\\[1mm]
              &\quad - \frac{2}{nm} \sum_{i=1}^{n}\sum_{j=1}^{m} k(x_i, y_j).
\end{align}

To derive a \textit{surrogate} version of the ELBO for the Quantum HyperNetworks problem, we replace the KL divergence term with the MMD term, scaled by a factor \( \lambda \):
\begin{align}
\mathcal{L}_{\text{SELBO}} = \mathbb{E}_{q_{\boldsymbol{\theta}}(\boldsymbol{\sigma})} 
\left[ \log p(Y \mid X, \boldsymbol{\sigma}) \right] %\notag \\ 
- \lambda \mathrm{MMD}^2(q_{\boldsymbol{\theta}}(\boldsymbol{\sigma}), p(\boldsymbol{\sigma})).
%\label{SurrogateELBO}
\end{align}
The scaling factor $\lambda$ determines the amount of regularization during optimization.
Different choices for $\lambda$ are possible, and the optimal one is highly problem-dependent. Annealing schedules are also possible \cite{blundell2015weight, Pomponi_2021}, to gradually transition towards \textit{Maximum Likelihood Estimation} (MLE) as training progresses. In our study, we consider $\lambda$ as a constant term, and we leverage the regularization term primarily to enhance the trainability of the Quantum HyperNetwork and to analyze its impact on the loss landscape. 

\begin{figure*}[tbh]
    \centering
    \includegraphics[width=0.96\linewidth]{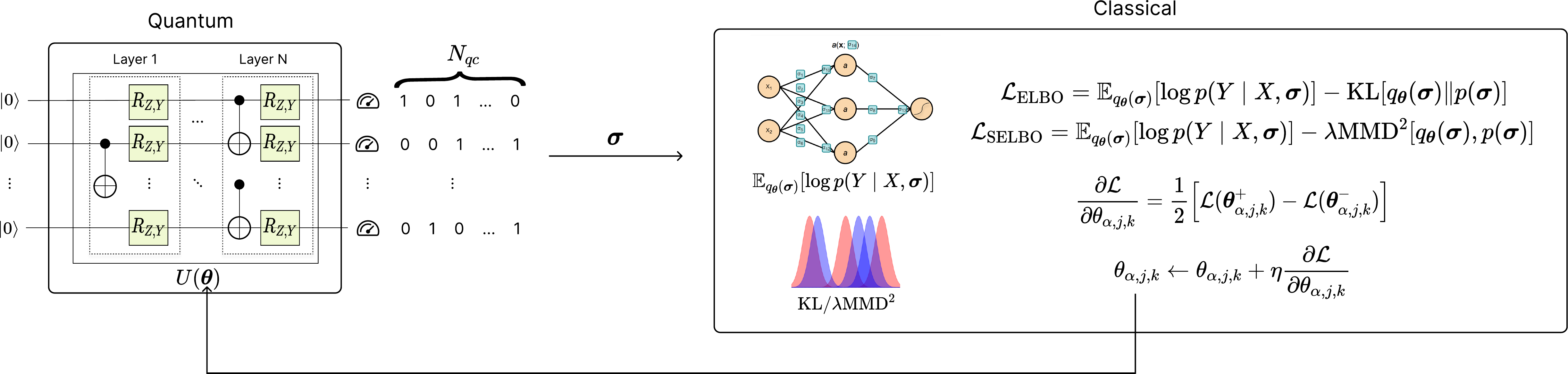}
    \caption{Representation of the Regularized Quantum HyperNetworks algorithm.}
    \label{fig:algorithm}
\end{figure*}

\section{Experiments}
\begin{figure*}[tbh]
    \centering
    \includegraphics[width=0.3\linewidth]{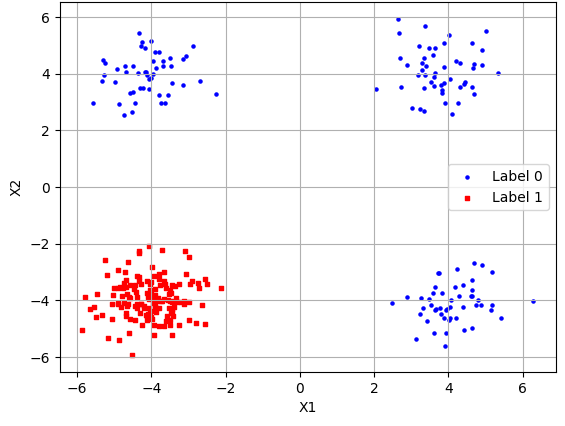}
    \includegraphics[width=0.3\linewidth]{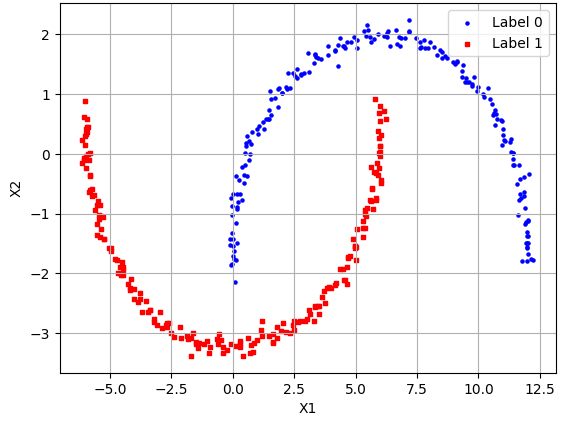}
    \includegraphics[width=0.23\linewidth]{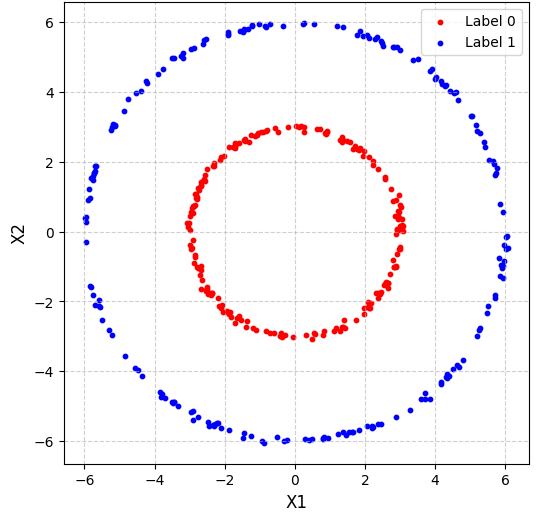}
    \caption{Datasets used in the experiments: 2D Gaussian dataset (a),  2D Moon dataset (b),  2D Ring dataset (c).}
    \label{fig:Datasets}
\end{figure*}

We evaluate our approach using three different toy datasets, illustrated in \cref{fig:Datasets}. The first dataset (a), taken from \cite{carrasquilla2023quantum}, consists of a two-dimensional distribution with four Gaussian clusters. 
The other two datasets, (b) and (c), correspond to 2D Moon and Ring shapes, respectively. All three datasets are balanced, with each of the two classes containing 150 data points for training and 100 samples for testing.
In all cases, we consider a binary classification problem, where the objective is to accurately predict the class of each 2D data point.

To perform classification, we use a simple BiNN consisting of a single hidden layer with 3 neurons and a total of 14 binary parameters, similarly to \cite{carrasquilla2023quantum}. Note that the value of the last bit in the measured bitstring, denoted as $\sigma_N$, determines the activation function for the hidden layer: if $\sigma_N=0$, ReLU is used, otherwise Sigmoid is applied.
The output layer uses a fixed Sigmoid activation function.

For the optimization task, we used gradient ascent with a variable learning rate, initially set to a high value of $\eta=1$: if the improvement in training loss falls below a threshold  $\min\delta$ for a given number of consecutive epochs (defined by the \textit{Patience} parameter), the learning rate is reduced by a predefined \textit{Decay Factor}. This adaptive process continues until the total number of epochs, $n_{\text{epochs}}$, is reached. A high initial learning rate is essential, as noted in \cite{wang2023bitnet}, since the low precision of binary network weights means that small parameter updates often yield negligible performance improvements.
All algorithmic specifications are detailed in \cref{tab:training_params}.

\begin{table}%[tbp]
\caption{Training Parameters}
\begin{center}
\begin{tabular}{|l|c|}
\hline
\textbf{Parameter} & \textbf{Value} \\
\hline
Number of Epochs ($n_{\text{epochs}}$) & 200 \\
Patience & 3 \\
Minimum Delta ($\min\delta$) & $10^{-4}$ \\
Decay Factor & 0.5 \\
Learning Rate ($\eta$) & 1 \\
Number of Keys ($n_{\text{keys}}$) & 100 \\
\hline
\end{tabular}
\label{tab:training_params}
\end{center}
\end{table}

We randomly initialize the gate parameters of the quantum circuit using a uniform distribution over the interval $[0,2\pi]$, and repeat the optimization process across $100$ different initializations to ensure statistical significance.

In our experiments using SELBO, we set the bandwidth parameter to \( h = \frac{n_{\text{qubits}}}{4} \), following the considerations outlined in \cite{rudolph2024trainability}. This choice links the bandwidth to the number of qubits in the system, facilitating comparisons within a continuous embedding space of discrete samples.
The underlying motivation is that the separation between samples is determined by the value of each qubit, which is 1, making this scaling choice essential for maintaining consistency across different configurations.

The procedure in \cref{alg:quantum_opt} is executed using the \textit{qujax} library \cite{qujax2023}, a quantum software framework designed for efficient quantum circuit simulation on GPUs. This simulation provides direct access to the full state-vector representation, including the corresponding amplitudes and probabilities of all possible outcomes.  
As a result, we can compute the entropy term in the explicit ELBO formulation without relying on repeated measurements to approximate the bitstring distribution, thus optimizing the loss function in \eqref{ELBO_approximate}.

\begin{algorithm}%[htb]
    \caption{Regularized Quantum HyperNetworks}
    \label{alg:quantum_opt}
    \begin{algorithmic}[1]
        \State \textbf{Input:} Initial quantum circuit parameters $\boldsymbol{\theta}$, number of epochs $n_{\text{epochs}}$, input data
        \State \textbf{Output:} Optimized parameters $\boldsymbol{\theta}^{\star}$

        \State $\boldsymbol{\theta}^{\star}\gets\boldsymbol{\theta}$,  $\mathcal{L}^\star \gets \Call{selbo}{\boldsymbol{\theta}}$
        \For{$i \gets 1$ to $n_{\text{epochs}}$}
            \State Compute gradients using the parameter-shift rule:
            \begin{equation*}
                \frac{\partial \mathcal{L}}{\partial \theta_{\alpha,j,k}} = \frac{\Call{selbo}{\boldsymbol{\theta}_{\alpha,j,k}^+} - \Call{selbo}{\boldsymbol{\theta}_{\alpha,j,k}^-}}{2}
            \end{equation*}
            \State Update parameters using gradient ascent:
            \begin{equation*}
                \theta_{\alpha,j,k} \gets \theta_{\alpha,j,k} + \eta \frac{\partial \mathcal{L}}{\partial \theta_{\alpha,j,k}}
            \end{equation*}
            \If {$\Call{selbo}{\boldsymbol{\theta}}
            > \mathcal{L}^\star$}
                \State $\boldsymbol{\theta}^{\star}\gets \left[\theta_{\alpha,j,k}\right]$, $\mathcal{L}^\star \gets \Call{selbo}{\boldsymbol{\theta}}$
            \EndIf

        \EndFor
        \State \Return{$\boldsymbol{\theta}^{\star}$}
    \Function{selbo}{$\boldsymbol{\theta}$}
    \State Measure $N_{\text{qc}}$ bitstrings from the quantum circuit  $|\Psi_{\boldsymbol{\theta}}\rangle$
    \State Upload the binary weights into the BiNN
    \State \textbf{return} $\mathcal{L}_\mathrm{(S)ELBO}(\boldsymbol{\theta})$ 
    \EndFunction
    \end{algorithmic}
\end{algorithm}

We compare the performance of our proposed methods, based on the ELBO and its implicit variant (SELBO), with results obtained using MLE, as employed in previous work \cite{naim2024scalable}. All experiments are conducted using quantum circuits with $N_\text{layers}={1}$ and a fixed number of measurements $N_{\text{qc}}=100$.
We opt for a shallow circuit depth because a low amount of entanglement is sufficient for the specific task at hand \cite{carrasquilla2023quantum}. In fact, the best solutions are often obtained with circuits containing fewer layers, which also makes them more amenable to classical simulation in small-scale settings.
Notice that the circuit parameters are initialized identically across all three methods to ensure a fair comparison.

\begin{figure*}[tbh]
\centerline{\includegraphics[width=0.9\textwidth]{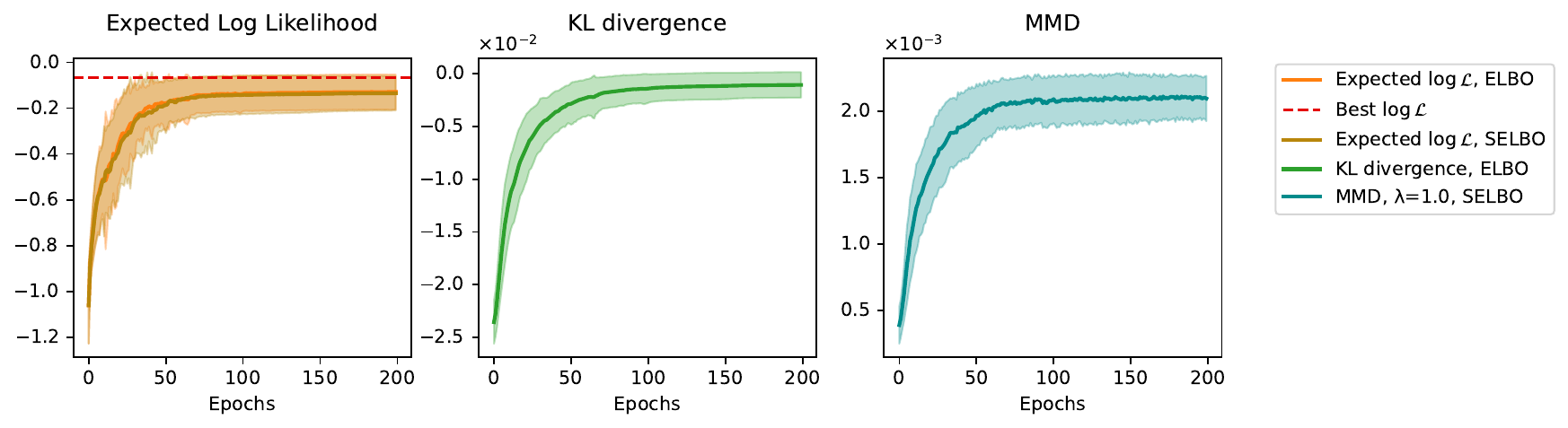}}
\caption{Average (S)ELBO for $N_\text{layers}=1$, $N_{\text{qc}} = 100$, run for 100 different initializations. Gaussian dataset. The KL is represented up to the constant term.}
\label{fig:(S)ELBO_result}
\end{figure*}

In \cref{fig:(S)ELBO_result}, we report the ELBO curves for the SELBO and ELBO methods on the Gaussian dataset. We represent the average behavior over the 100 initializations (random seeds), where we show the mean value and the standard deviation over the different initializations. We can see how the expected log-likelihood and the regularization term are correctly increasing, both in the explicit and implicit cases, meaning that we are moving away from the uniform prior. 

\begin{figure*}[tbh]
\centerline{\includegraphics[width=1\textwidth]{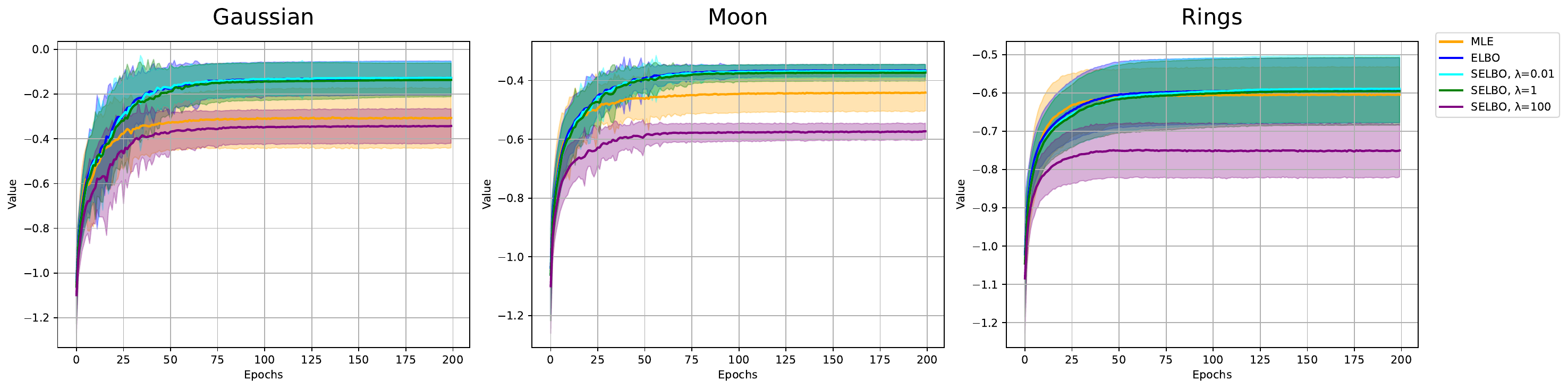}}
\caption{Training curves for (S)ELBO and MLE, $N_\text{layers}=1$, $N_\text{qc} = 100$, 100 different initializations.}
    \label{fig:comparison_1_100}
\end{figure*}

In \cref{fig:comparison_1_100} we show the training curves for different methods (ELBO, SELBO and MLE) across the three datasets. We can see that the (S)ELBO values for $\lambda \in \{0.01,1\}$ go above the MLE in spite of the penalization, thus indicating a better trainability.
Moreover, it seems to help for the optimization task, intuitively smoothing the high-dimensional search space and helping it to escape from local optima. Note that the amount of regularization is also an important factor: considering high value of $\lambda$ the SELBO diverges from the original ELBO, possibly degrading the performance. However, from the experimental results, we notice that there is a range of $\lambda$ values that can be considered to achieve good performance for the classification task, whereas for higher values (i.e. from $\lambda=1000$) the regularization overcomes the expected log-likelihood, leading to bad binary configurations.

In \cref{tab:merged_losses}-\cref{tab:merged_accuracies}, we report the final \textit{binary cross entropy} (BCE) losses and accuracies for each dataset on the test set, trained with the three different methods. Note that, for the SELBO, we consider a constant $\lambda \in \{0.01,1.0,100.0\}$. The ``MLE E.S.'' in the tables refers to the test loss considering the best binary configuration on the training set, trained with MLE via Exhaustive Search, that is the brute force approach trying all the possible $2^{N_\text{qubits}}$ configurations. ``MLE'' in the tables refers to the quantum circuit approach of \cite{naim2024scalable}, using gradient ascent and Monte Carlo sampling.

On average, the SELBO outperforms the standard MLE method. We observe that, in most cases, the (S)ELBO leads to a higher and more concentrated accuracy. The best solution is found with the SELBO method and $\lambda=0.01$, meaning that only a small amount of regularization is needed to achieve better solutions.
However, we note that for the Rings dataset, the problem is generally more challenging to solve due to the limitations of the simple binary neural network under study.

We also acknowledge that, thanks to the regularization term in the VI formulation, one can also reach better performances than those obtained with ``MLE E.S.''. This is reasonable because the optimal solution on the training set may not necessarily be the best on the test set. Instead, the Quantum HyperNetworks approach overcomes this issue by leveraging the uncertainty in sampling binary configurations: this is achieved through the probabilistic distribution, which allows for a more robust exploration of possible configurations, improving generalization beyond the training set.

\begin{table*}%[tbp]
\caption{Final BCE loss comparison between MLE, ELBO, and SELBO on the test sets. The lower the better.}
\begin{center}
\begin{tabular}{|l|c|c|c|c|c|c|}
\hline
\textbf{Dataset} & \textbf{MLE E.S.} & \textbf{MLE} & \textbf{ELBO} & \textbf{SELBO (\(\lambda=0.01\))} & \textbf{SELBO (\(\lambda=1.0\))} & \textbf{SELBO (\(\lambda=100.0\))} \\
\hline
GAUSSIAN & 0.067 & 0.315 $\pm$ 0.123 & 0.139 $\pm$ 0.073 & \textbf{0.129 $\pm$ 0.071} & 0.137 $\pm$ 0.074 & 0.141 $\pm$ 0.083 \\
\hline
MOON     & 0.357 & 0.450 $\pm$ 0.062 & 0.378 $\pm$ 0.024 & \textbf{0.371 $\pm$ 0.022} & 0.374 $\pm$ 0.027 & 0.387 $\pm$ 0.033 \\
\hline
RINGS    & 0.520 & 0.612 $\pm$ 0.070 & 0.608 $\pm$ 0.085 & \textbf{0.599 $\pm$ 0.084} & 0.603 $\pm$ 0.086 & 0.622 $\pm$ 0.085 \\
\hline
\end{tabular}
\label{tab:merged_losses}
\end{center}
\end{table*}

\begin{table*}%[tbp]
\caption{Accuracy Comparison Between MLE, ELBO, and SELBO on the test sets. The higher the better.}
\begin{center}
\begin{tabular}{|l|c|c|c|c|c|c|}
\hline
\textbf{Dataset} & \textbf{MLE E.S.} & \textbf{MLE} & \textbf{ELBO} & \textbf{SELBO (\(\lambda=0.01\))} & \textbf{SELBO (\(\lambda=1.0\))} & \textbf{SELBO (\(\lambda=100.0\))} \\
\hline
GAUSSIAN & 0.993 & 0.893 $\pm$ 0.108 & 0.986 $\pm$ 0.012 & \textbf{0.991 $\pm$ 0.011} & 0.990 $\pm$ 0.012 & 0.987 $\pm$ 0.019 \\
\hline
MOON     & 0.810 & 0.766 $\pm$ 0.082 & 0.808 $\pm$ 0.008 & \textbf{0.810 $\pm$ 0.006} & 0.809 $\pm$ 0.004 & 0.806 $\pm$ 0.008 \\
\hline
RINGS    & 0.855 & 0.698 $\pm$ 0.101 & 0.698 $\pm$ 0.141 & \textbf{0.712 $\pm$ 0.145} & 0.707 $\pm$ 0.144 & 0.671 $\pm$ 0.137 \\
\hline
\end{tabular}
\label{tab:merged_accuracies}
\end{center}
\end{table*}

\begin{figure*}[h!]
\centerline{\includegraphics[width=\textwidth]{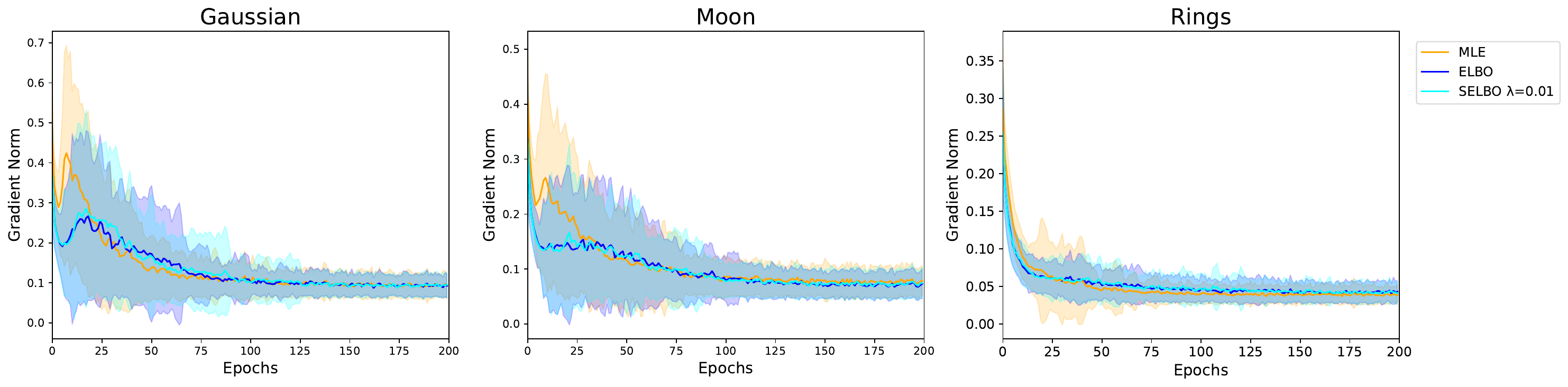}}
\caption{Evolution of the average gradient magnitude during training.}
\label{fig:gradients}
\end{figure*}

Finally, to better understand the effect of regularization on the optimization process, we plot the evolution of the average gradient magnitude during training for the three different datasets in \cref{fig:gradients}. 

One can see that, when the training curves in \cref{fig:comparison_1_100} start to diverge, the gradients for the MLE case are higher and less concentrated than those employing the (S)ELBO. Additionally, one can note that for the Rings dataset, even if the problem is hard to solve for the simple BiNN under consideration, the (S)ELBO gradients are still slightly better than those obtained with MLE.

\begin{figure*}[h!]
\centerline{\includegraphics[width=0.9\textwidth]{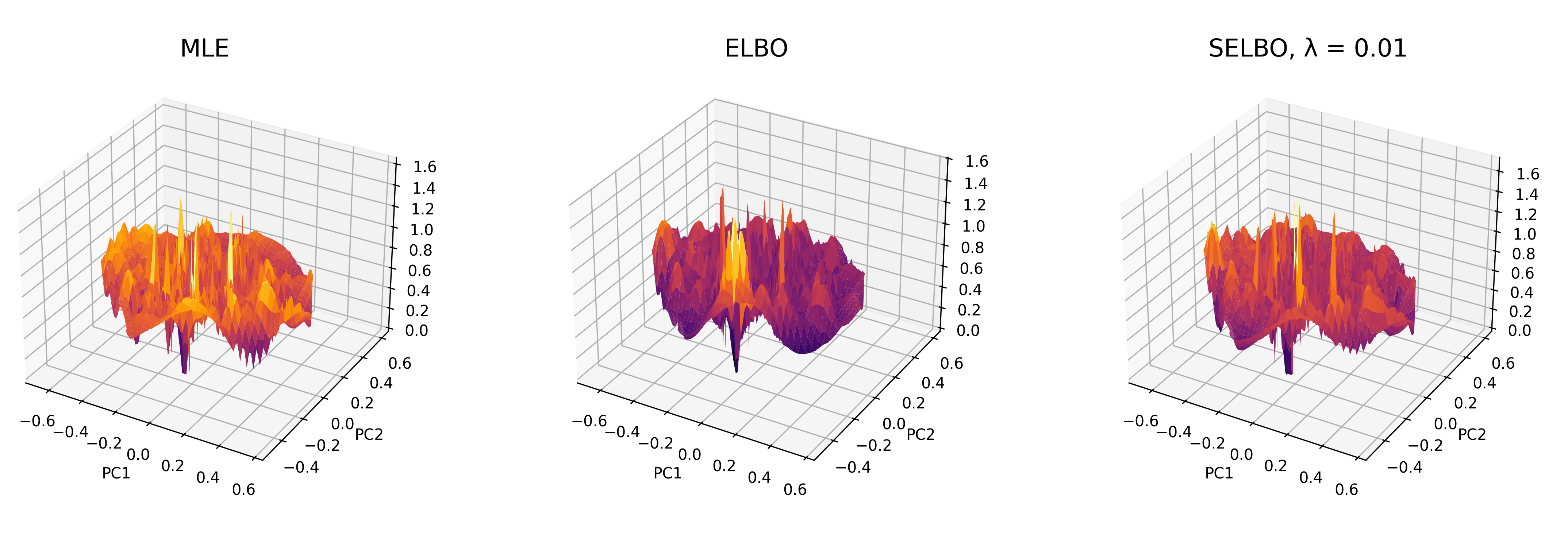}}
\caption{Loss landscapes for different optimization procedures. Gaussian dataset.}
\label{fig:LossLandscapes}
\end{figure*}

We also illustrate in
\cref{fig:LossLandscapes} the loss landscapes around a locally optimal point for the Gaussian dataset, comparing the three different methods. It can be observed that regularization smooths the landscape, thereby facilitating optimization and enhancing trainability.

Regarding the necessary computing resources to run the experiments, a typical run on these toy datasets takes about 55 minutes on an NVIDIA Tesla T4 GPU, consuming about 150 kJ with approximately 12 GB of peak GPU memory usage. Given these high computing requirements, it is challenging to perform simulations with larger datasets and networks.

\section{Conclusions and future work}
In this work, we established a connection between \textit{Quantum HyperNetworks} and Bayesian inference by deriving the \textit{Evidence Lower Bound} (ELBO) for optimizing \textit{Binary Neural Networks} using quantum circuits. We introduced two formulations: an \textit{explicit} ELBO, which leverages direct access to the quantum circuit distribution, and a \textit{surrogate} ELBO, tailored for scenarios where only implicit distributions are accessible, as commonly encountered in practical quantum hardware implementations.  

Our results show that optimizing BiNNs with ELBO-based variational methods enhances both trainability and generalization compared to conventional Maximum Likelihood Estimation. The ELBO framework facilitates better parameter updates by smoothing the loss landscape, leading to more efficient and generalized training. Moreover, this method could mitigate possible overfitting issues in complex training scenarios, thanks to the employment of the regularization term.

Our findings underscore the benefits of using an ELBO-driven approach for BiNN training and suggest that quantum-inspired variational methods may offer a principled pathway toward more effective learning. Future work could further explore the theoretical relationship between explicit and surrogate ELBO formulations, particularly through the KALE divergence \cite{glaser2021kale}. Additionally, improvements in parameter initialization for quantum circuits could improve convergence stability and training efficiency. Empirical validation on real quantum hardware will be crucial in assessing the practical feasibility of the proposed approach, particularly in a noisy environment. Furthermore, investigating the scalability of this method, especially in the context of \textit{Multi-Basis Encoding} \cite{PhysRevResearch.4.033142, naim2024scalable}, will be essential for larger networks, as well as to extend this framework to support higher precision weights \cite{wang2025optimizing}. Lastly, since training multiple binary weights using a single qubit can potentially introduce performance degradation, developing effective mitigation strategies will be key to ensuring robust performance at larger scales.

%\section*{Acknowledgment}
%Write Acknowledgment.

%\printbibliography
%\vspace{12pt}

\bibliographystyle{IEEEtran}
\bibliography{IEEEabrv,bibliography}
\end{document}